\documentclass{emulateapj}

\newcommand{\lsim}{\lesssim}
\newcommand{\gsim}{\gtrsim}

\begin{document}

\title{Impact of Dark Matter Annihilation on the High-Redshift Intergalactic Medium}
\author{Leonid Chuzhoy\altaffilmark{1}}

\altaffiltext{1}{Department of Astronomy and Astrophysics, The University of Chicago, 5640 S. Ellis, Chicago, IL 60637, USA}

\begin{abstract}
We reexamine the impact of dark matter (DM) annihilation on the intergalactic medium, taking into account the clumping of DM particles. 
We find that energy injection from the annihilation of the thermal relic DM particles may significantly raise the gas temperature at high redshifts and leave a strong imprint on the cosmological 21-cm signal ($\delta T_b>10$ mK), provided the particle mass is below $\sim 1$ TeV.
Further, we find that while the energy injection from DM annihilation could not alone complete the reionization of the Universe, it could make a significant contribution to the electron optical depth.
\end{abstract}


\section{\label{Int}Introduction}

In the standard cosmological paradigm, the non-baryonic dark matter (DM) particles make above 80\% of the present matter density \cite{Sp}. However, despite their high abundance, the DM particles have so far successfully resisted all efforts to detect them. While this non-detection has produced strong constraints on the nature of the DM particle and the strength of its interaction with the ordinary matter, there is still enough room for the non-gravitational processes involving DM to make a strong impact on the observable Universe \cite{BS,QW,HO,CN,BY,Asc,MN}.

In this paper, we continue to explore the impact of the DM annihilation on the reionization history of the Universe. The products of the DM annihilation, which may include energetic photons and electron/positron pairs, are expected to couple to the ordinary matter, thereby raising the IGM temperature and the ionized fraction. In several previous papers, where this process was examined, it has been concluded that the effect of DM annihilation on the evolution of the intergalactic medium (IGM) during reionization should be rather minor \cite{MFP,FOP,RMF,RMFb,Val}. 
The main reason for this conclusion can be briefly described as following. Since the annihilation rate of non-relativistic DM particles is expected to depend primarily on their density, it should be the highest in the early universe, declining as $(1+z)^3$ as the universe expands. Consequently, for the annihilation rate, which is low enough not to disturb recombination of Universe at $z\sim 1000$ violating the constraints set by the WMAP \cite{Zal},  the energy input from DM annihilation during the epoch of reionization ($z\sim 10$) would be too low to make a significant impact.

However, as we show in this paper, this conclusion may have to be changed, once the DM clumping is taken into account. While the DM density field is expected to be nearly homogeneous prior and during the epoch of recombination, eventually the growth of density perturbation leads to formation of DM halos with constantly increasing density contrast. Consequently the DM annihilation rate 
declines much slower than the other processes that affect the IGM, such as inverse Compton (IC) scatterings and recombinations, and within the appropriate range of parameters DM annihilation can make a major contribution to reionization, even if it were relatively unimportant earlier.

In \S 2, we describe the evolution of the DM clumping. In \S 3 and 4, we estimate the impact of DM annihilation on the IGM and the cosmological 21-cm signal. We summarize our results in \S 5.

\section{DM clumping}
The evolution of the DM distribution can be qualitatively divided into several stages. In the early universe ($z\ll 100$) the DM density perturbations are generally expected to be small and the clumping factor, $F_{cl}=\langle\rho_X^2\rangle/\langle\rho_X\rangle^2$, close to unity. 
After a period of slow linear growth, density perturbations enter the non-linear regime, which is followed by formation of the DM halos.
Unfortunately, the timing of this transition is uncertain. The amplitude of the primordial small-scale density perturbations, which are first to enter the non-linear regime, cannot yet be probed observationally and may be estimated only by extrapolation from much higher scales. Further, the free-streaming of DM particles dilutes the perturbations by a factor that depends on a particle mass, which itself is an unknown quantity. Assuming the spectral index $n=1$, which is favored by the inflation models, the non-linear regime should start around $z\sim 50-70$ \cite{GHS}.  However, one can not yet exclude the possibility that on small scales the primordial perturbation spectrum is drastically different from the assumed power law. If so, the non-linear regime may start shortly before reionization, or, at the other extreme, the small-scale perturbations may be already non-linear at the recombination epoch. 

After microhalos undergo the process of virialization, their internal density evolution slows down and their density contrast with respect to the average density of the expanding universe increases as $\sim (1+z)^{-3}$. However, when microhalos merge into larger objects, their internal density is reduced by tidal forces and some microhalos may be completely evaporated. Numerical simulations indicate that after the merger the smaller halos typically lose most of their mass to the tidal forces of their neighbors or of the main halo. Nevertheless, the central regions of the microhalos remain relatively intact so that on average the tidal forces reduce the dark matter clumping is only by a factor of a few \cite{Kal,DKM,Gal,BDE}. Later the formation of stellar disks may result in almost complete destruction of microhalos \cite{AZ,Zhao}, but, since stars are expected to form in high numbers only at low redshifts, this is likely to happen only after the end of reionization.

\begin{figure}[t]
\resizebox{\columnwidth}{!} {\includegraphics{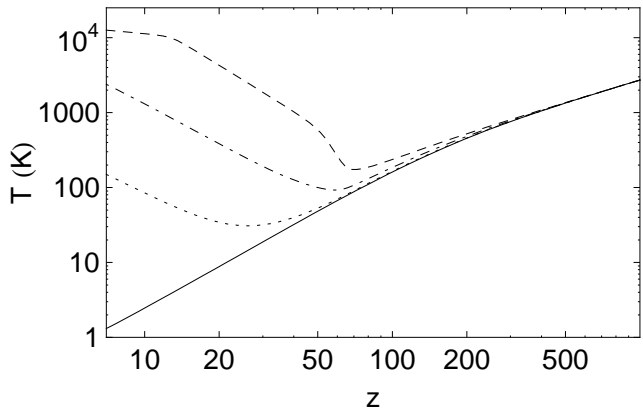}}
\caption{Thermal evolution of the IGM for $\zeta=$ $0$, $0.1$, $1$ and $10$  (solid, dotted, dot-dashed and dashed curves, respectively).} \label{Tev}
\end{figure}

\section{IGM evolution}
The energy input rate from DM annihilation is given by
\begin{equation}
 \label{rix}
I_{x}=gf_{abs}F_{cl}\left(\frac{\langle\rho_X\rangle}{m_X}\right)^2\langle\sigma v\rangle m_X c^2,
\end{equation}
where $f_{abs}$ is the fraction of the released energy absorbed by the IGM, $\langle\rho_X\rangle$ is the average DM density and $m_X$ is the mass of a single dark matter particle. It is generally expected that the annihilation cross-section of non-relativistic DM particles scales inversely with their velocity, so that their product $\langle\sigma v\rangle$ is a constant. The degeneracy factor $g$ equals $1$, if the DM particle are identical (i.e., the particle and the anti-particles are the same) or $1/2$ otherwise.
The value of $f_{abs}$ depends on the model for the DM particle.
Further, as shown by Ripamonti et al. (2007), it is likely to vary with redshift, possibly changing by more than an order of magnitude between $z=10$ and 1000. However, to avoid restricting the calculations to a particular DM model, we shall have to assume that $f_{abs}$ is a constant.

We assume that, as is the case with most DM particle models, the energy from DM annihilation is injected into the baryonic matter primarily through collisions with high-energy electrons, which may be either the direct product of the DM particles annihilation or produced by secondary ionizations.  A fraction of the absorbed energy is converted into heat, $f_{Heat}\approx(1-(1-x^{0.266})^{1.316})$, where $x$ is the ionized fraction, and the remainder is almost equally split between secondary ionizations and atomic excitations \cite{SvS}.

In the following calculations we assume that after entering the non-linear phase at $z\sim 60$, the DM clumping grows as $F_{cl}\propto (1+z)^{-3}$. Prior to that epoch, we make a simplistic assumption that the density perturbations evolve as spherical top-hats. It should be noted though that the state of the IGM during reionization is almost insensitive to the assumptions we make for $z\gsim 60$, since most of the energy from the DM annihilation would be released long after that epoch. 

The evolution of the IGM depends on the product of several uncertain parameters
\begin{equation}
\zeta=\left(\frac{f_{abs}}{0.1}\right) \left(\frac{\langle\sigma v\rangle}{10^{-26}{\rm cm^{-3}\; s^{-1}}}\right) \left(\frac{F_{cl}(1+z)^{3}}{10^8}\right)  \left( \frac{m_X}{\rm GeV}\right)^{-1}. 
\end{equation}
To set a plausible range for the value of $\zeta$, we have to consider separately the plausible values for $f_{abs}$, $\langle\sigma v\rangle$ and $F_{cl}(1+z)^{3}$.
For $n=1$,
$F_{cl}(1+z)^{-3}$ should of order $5\cdot 10^7$ (for comparison, $F_{cl}(1+z)^{-3}$ of the DM halo that forms at $z=30$ is $\sim 6\cdot 10^7$, assuming the NFW density profile and the concentration parameter c=4 that is characteristic of the newly formed halos).
If, as usually assumed, the DM particles were in thermal equilibrium during their decoupling epoch, their annihilation cross-section is around 
$\langle\sigma v\rangle\sim 2\cdot 10^{-26}\; {\rm cm^{-3}\; s^{-1}}$.
During the epoch of reionization $f_{abs}$ is likely to be of order $0.1$ \cite{RMF}. Thus we expect $\zeta$ to be of order $(m_X/{\rm GeV})^{-1}$. 

The WMAP observations set an additional constraint on the DM annihilation rate \cite{Zal}
\begin{equation}
\langle \sigma v \rangle\lsim 0.8\cdot 10^{-26}f_{abs}^{-1} \left(\frac{m_X}{\rm GeV}\right)\; {\rm cm^{-3}\; s^{-1}},
\end{equation}
which for our choice of $F_{cl}$ translates into $\zeta\lsim 4$.

In addition to the energy injection from DM annihilation, the thermal evolution of the IGM is set by the adiabatic cooling, the change in the ionization state, the inverse Compton (IC) scattering of cosmic microwave background (CMB) photons and the radiative cooling processes
\begin{equation}
\frac{dT}{dt}=\frac{2f_{Heat}I_x-\L_{rad}}{3k_Bn(1+x)}-\frac{4T}{3t}-\frac{Tdx}{1+x}+\frac{x}{1+x}\frac{T_{cmb}-T}{t_\gamma},
\end{equation}
where $T$ is the IGM temperature, $t_\gamma\propto (1+z)^{-4}$ is the IC cooling time-scale, $k_B$ is the Boltzmann constant, $n$ is the baryon number density and $T_{CMB}$ is the temperature of the CMB photons. The radiative cooling, $L_{rad}$, is dominated by collisional excitation of hydrogen atoms and hydrogen recombinations (we used the rates from Dalgarno \& McCray (1972) and Seaton (1959)).
The evolution of ionized fraction is dominated by radiative recombination and secondary ionizations induced by DM annihilation
\begin{equation}
\frac{dx}{dt}=-\alpha(T)x^2 n+\frac{f_{ion}I_x}{nE_H},
\end{equation}
where $\alpha(T)$ is the recombination rate, $E_H=13.6$ eV is the hydrogen ionization potential and $f_{ion}\approx(1-f_{Heat})/2$ is the fraction of the absorbed energy that goes into ionizations.

\begin{figure}[t]
\resizebox{\columnwidth}{!} {\includegraphics{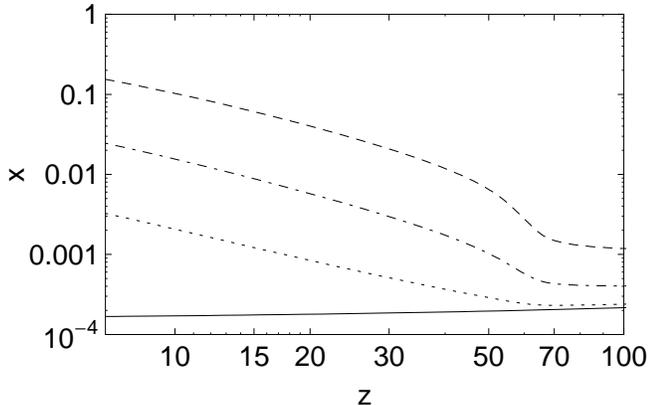}}
\caption{Same as Fig. \ref{Tev} for ionization state.} \label{xev}
\end{figure}

\begin{figure}[t]
\resizebox{\columnwidth}{!} {\includegraphics{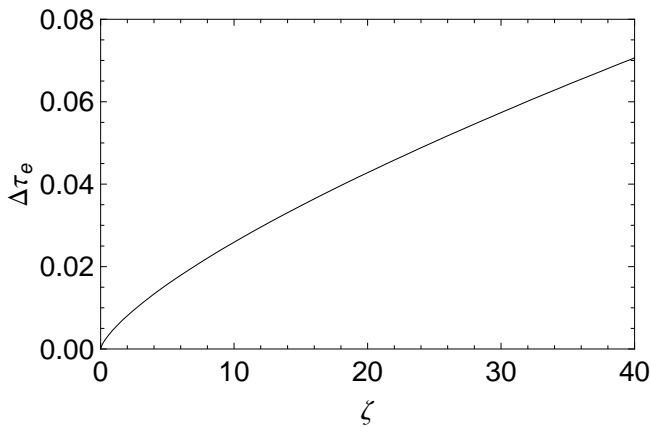}}
\caption{The DM annihilation contribution to the electron optical depth.}\label{tau}
\end{figure}

Figures \ref{Tev} and \ref{xev} show the impact of the DM annihilation on thermal evolution and the ionization state of the IGM for different values of $\zeta$.
Figure \ref{tau} shows the electron optical depth, $\Delta \tau_e$, for the redshift interval $7<z<150$, assuming the absence of other ionizing sources. When $\zeta$ is close to its upper limit, we see that $\Delta\tau_e $ can exceed 0.01. Further, if the DM clumping is a few times higher than we assumed, which is conceivable due to a large uncertainty in the small-scale DM distribution, then the $\Delta \tau_e$ can make a large fraction of the total electron optical depth, $\tau_e\approx 0.09$ \cite{WMAP}. 
However, we find that DM annihilation cannot alone complete the reionization of the Universe. Since energy injection from the DM annihilation changes rather slowly, the end of reionization ($x=1$) must be preceded by a long period of significant partial ionization. Consequently, the energy injection rate sufficient to complete reionization by $z\sim 7$ results in the total electron optical depth being several times higher than allowed by the WMAP observations.

\section{The 21-cm signature}
The next generation of radio-telescopes (LOFAR, MWA, SKA) should be capable of detecting the redshifted 21-cm continuum from the epoch of reionization. Since the 21-cm signal is sensitive to the gas temperature, ionized fraction and the UV intensity around the Ly$\alpha$ resonance, DM annihilation, which affects each of these parameters, can produce a detectable imprint on the observed signal.
To illustrate this, we calculate the evolution of the 21-cm spin and the differential brightness temperature, $T_s$ and $\delta T_b$.

The spin temperature is a weighted function of the CMB temperature, $T_{\rm CMB}$,
and the kinetic temperature of hydrogen, $T_k$,  \citep{f8}
\begin{equation}
T_s=\frac{T_{\rm CMB}+y_\alpha T_k+y_cT_k}{1+y_\alpha+y_c},
\end{equation}
where $y_c$  is the collisional coupling constant (see Zygelman (2005)
and Liszt (2001) for the contribution of neutral atoms and electrons to the collisional pumping). The radiative coupling constant,
$y_\alpha$, is 
\begin{equation}
\label{eqn:y}
y_\alpha=\frac{16\pi^2 T_*e^2 f_{12}J_0}{27 A_{10}T_k m_e c}S_\alpha,
\end{equation}
where $f_{12}=0.4162$ is the oscillator strength of the Ly$\alpha$
transition, $T_*=h\nu_*/k=0.068$ K, $A_{10}$ is the spontaneous
emission coefficient of the hyperfine transition, $m_e$ and $e$ are electron mass and charge, and $J_0$ is the
intensity at Ly$\alpha$ resonance, when the backreaction on the
incident photons caused by resonant scattering is neglected \cite{f8,CM,Hir,CS6,FP}. For the
unperturbed IGM the backreaction correction, $S_\alpha$, is
\citep{CS6}
\begin{equation}
\label{back}
S_\alpha=e^{-0.37(1+z)^{1/2}T_k^{-2/3}}\left(1+\frac{0.4}{T_k}\right)^{-1}.
\end{equation}
The Ly$\alpha$ resonance photons can be produced by the conventional UV sources (i.e., stars and AGNs) or indirectly by DM annihilation. \footnote{The scatterings of non-thermal electrons, produced by DM annihilation, with hydrogen atoms, result in electron excitations, which in $\sim 5$ out of 6 cases are followed by the emission of the Ly$\alpha$ photon\cite{CAS}.}

For unperturbed IGM the differential brightness temperature presently
observed at the wavelength $21(1+z)$ cm is \cite{MMR}
\begin{eqnarray}
\delta T_{\rm b}\approx 0.03\; {\rm K} \left(\frac{T_{\rm s}-T_{\rm CMB}}{T_{\rm s}}\right) (1-x)  \left(\frac{1+z}{10}\right)^{1/2}.
\end{eqnarray}

\begin{figure}[t]
\resizebox{\columnwidth}{!} {\includegraphics{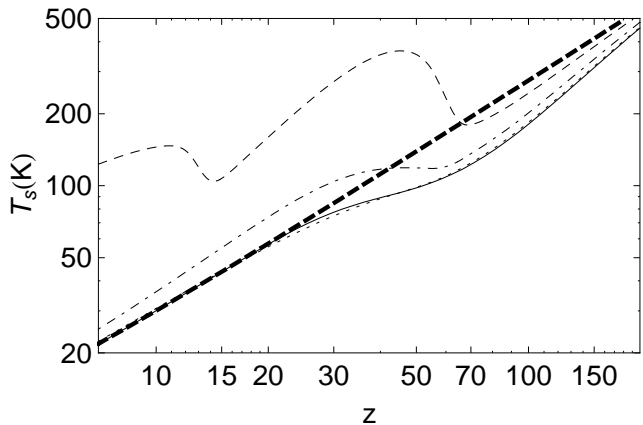}}
\caption{Same as Fig. \ref{Tev}  for the 21-cm spin temperature. The thick dashed line shows the CMB temperature.} \label{tsf}
\end{figure}

\begin{figure}[t]
\resizebox{\columnwidth}{!} {\includegraphics{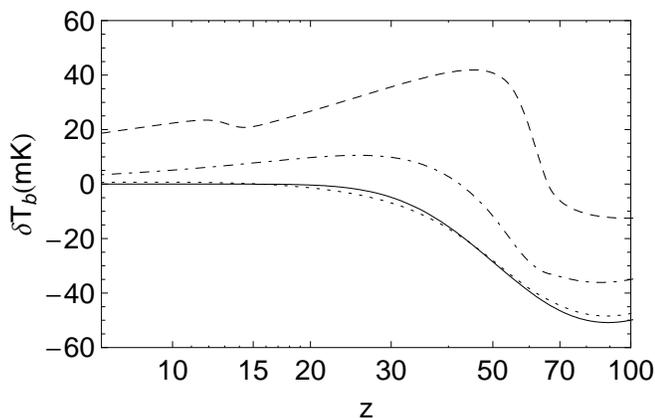}}
\caption{Same as Fig. \ref{Tev}  for the  21-cm differential brightness temperature.} \label{tbf}
\end{figure}

Figures \ref{tsf} and \ref{tbf} illustrate the evolution of $T_s$ and $T_b$ in the absence of all other radiation sources. When the energy input from DM annihilation is relatively high ($\zeta\gsim 1$), the evolution of $T_s$ can be distinctly different from other reionization scenarios. For instance, in models where stars and miniquasars are the only radiation sources available, the decrease of $T_s$ due to the expansion of the Universe is expected to be reversed only a short time before reionization, at $z\lsim 20$. By contrast, the DM annihilation may cause the decrease of $T_s$ be reversed first already at $z\sim 50$, when the DM clumping enters the phase of fast non-linear growth. Subsequently $T_s$ can decline again until Ly$\alpha$ photons are produced in large number either by stars or the line cooling in the IGM.

Even a relatively low energy injection rate ($\zeta<0.01$) may produce an observable effect. In this case the DM annihilation does not create sufficient numbers of the Ly$\alpha$ photons to decouple $T_s$ from $T_{CMB}$ on its own. However, it can still produce a noticeable effect on the gas kinetic temperature (Fig. \ref{Tev}) and, when the UV background from other sources becomes sufficiently high, this would be reflected in the 21-cm signal. In this way, raising the gas temperature by just a few degrees at $z\sim 12$, might change $\delta T_b$ by more than 100 mK, which is above the expected sensitivity ($\sim 10$ mK) of LOFAR.

\section{Discussion}
The high value of reionization optical depth, $\tau_e$, measured by WMAP, indicates that reionization might begin significantly earlier than the observations of the present-day galaxies would suggest, and that new types of radiation sources may be required to explain it \cite{Gn}. The high opacity of the IGM to the Ly$\alpha$ photons at $z\gsim 6$ further indicates that reionization proceeded over very extended period. Thus preionization produced by DM annihilation can fit well into the existing constraints on the reionization scenarios. 
Naturally, not all of the DM candidates are suitable. The constraint on the relic abundance of the thermalized DM particles, $\langle\sigma v\rangle\lsim 2\cdot 10^{-26}\; {\rm cm^{-3}\; s^{-1}}$, rules out most of the heavier DM candidates, such as the neutralino, whose annihilation rate would be too low. On the other hand, some of the lighter DM candidates with mass below 100 MeV \cite{Hal} may fit the profile.

DM particles, whose annihilation rate is too low to make a significant impact on the reionization, might still leave a potentially detectable signature in the cosmological 21-cm background, which is sensitive to the gas temperature.
However, the heat input from DM annihilation may turn out to be secondary to other processes, such as X-ray absorption or scatterings of UV resonance photons \cite{Ven,OH,CS7,CiS}. In addition, given that during the reionization epoch the DM particles annihilation time-scale is expected to remain nearly constant, the effect of the DM annihilation may be difficult to distinguish from the effect of DM decay.

\end{document}